# Enhanced Single Photon Emission from a Diamond-Silver Aperture

Jennifer T. Choy[1*], Birgit J. M. Hausmann[1*], Thomas M. Babinec[1*], Irfan Bulu[1*], Mughees Khan[2], Patrick Maletinsky[3], Amir Yacoby[3], Marko Lončar[1+]

[1] School of Engineering and Applied Sciences, Harvard University, Cambridge, MA 02138, USA

[2] Wyss Institute, Harvard University, Cambridge, MA 02138, USA

[3] Department of Physics, Harvard University, Cambridge, MA 02138, USA

[*] These authors contributed equally to this work.

[+] Corresponding author: loncar@seas.harvard.edu

## Abstract

We have developed a scalable method for coupling single color centers in diamond to plasmonic resonators and demonstrated Purcell enhancement of the single photon emission rate of nitrogen-vacancy (NV) centers. Our structures consist of single nitrogen-vacancy (NV) center-containing diamond nanoposts embedded in a thin silver film. We have utilized the strong plasmon resonances in the diamond-silver apertures to enhance the spontaneous emission of the enclosed dipole. The devices were realized by a combination of ion implantation and top-down nanofabrication techniques, which have enabled deterministic coupling between single NV centers and the plasmonic modes for multiple devices in parallel. The plasmon-enhanced NV centers exhibited over six-fold improvements in spontaneous emission rate in comparison to bare nanoposts and up to a factor of 3.6 in radiative lifetime reduction over bulk samples, with comparable increases in photon counts. The hybrid diamond-plasmon system presented here could provide a stable platform for the implementation of diamond-based quantum information processing and magnetometry schemes.

Solid-state single photon sources, such as the nitrogen-vacancy (NV) center in diamond[1], are robust systems for practical realizations of various quantum information processing protocols[2-8] and nanoscale magnetometry schemes[9-10] at room temperature. Such applications benefit from high emission efficiency and flux of single photons, which can be achieved by engineering the emitter's electromagnetic environment. One particularly attractive approach is based on plasmonic resonators[11-16], in which sub-wavelength confinement of optical fields can strongly modify the spontaneous emission of a suitably embedded dipole. Meanwhile, the scalability of solid-state quantum systems critically depends on the ability to control such emitter-cavity interaction in a number of devices in parallel. Here, we demonstrate a scalable method to enhance the radiative emission rate of single NV centers using plasmonic apertures that preserve the critical properties of the NV center for quantum information, including nonclassical photon statistics and optically-detected electron spin resonance contrast.

Efficient single photon-generation and -extraction from solid-state quantum emitters is an important problem in quantum photonic devices and systems. In the case of NV center-based devices in diamond, the photon generation rate and outcoupling efficiency are limited by the relatively long radiative lifetime (~12ns) and total internal reflection (TIR) at the diamond-air interface. Approaches to overcome this problem in the past have included direct fabrication of nanowires[17] and solid-immersion-lenses[18] in bulk diamond crystals, as well as the use of diamond nanocrystals[19] coupled to dielectric[20-24] or metallic[25] optical resonators.

The bottleneck in many of these techniques is the deterministic coupling of single quantum emitters to photonic elements, which is typically challenging and incompatible with high throughput, large-scale production of devices. In this work, we present an approach to overcome this limitation by directly embedding single NV centers into metallic nanostructures.

Our devices support highly localized optical fields which are designed to provide a high yield in coupling to the NV center fluorescence (Fig. 1(a)). Specifically, we consider plasmonic apertures, consisting of cylindrical diamond nanoposts (radius r~50nm, height h~180nm) surrounded by silver (Ag). These structures support modes with mode volumes as small as $0.07(\lambda/n)^3$ and can provide good spatial overlap between the strong optical fields and enclosed dipole due to nearly uniform field distributions in the transverse direction (Figs. 1(b)-(c)). This results in the enhancement of the spontaneous emission (SE) rate of the dipole.

The SE rate enhancements for our structures were calculated with a three-dimensional finite-difference-time-domain (3D FDTD) solver using a classical approach[16] by comparing the total power emitted from a dipole when it is placed inside the aperture to the total power emitted in a homogeneous medium. The theoretical SE rate enhancement spectrum, plotted in Fig. 1(d) for aperture radii of 50nm, 55nm, and 65nm, exhibits a broad resonance that red-shifts with increasing radius and can therefore be tailored to overlap with the NV emission while keeping the height of the structure constant[16]. Based on our simulations, SE rate enhancements on the order of ~30 can be expected for an in-plane polarized NV center placed at the maximum field intensity in an optimized structure.

The hybrid diamond-metal device, depicted in Fig. 1(a), was realized using a combination of blanket ion implantation and top-down nanofabrication techniques[17,26-27] (Fig. 2(a)). An ultrapure bulk diamond crystal (type IIa, Element 6) was implanted with nitrogen ions and subsequently annealed[27] to generate a layer of NV centers approximately 20nm below the diamond surface. Arrays of diamond nanoposts of radii from 50nm to 65nm and height~180nm were then fabricated using electron beam lithography followed by inductively coupled plasma reactive ion etching (ICP RIE) (Fig. 2(b)). We have previously shown[27] that such a procedure can result in a high yield (>10%) of single-center devices. The nanoposts were finally embedded in a 500nm-thick Ag film that was deposited by electron beam evaporation. Details of the fabrication procedure are further elaborated in the Methods section.

In order to rigorously quantify the effect of the plasmon cavity on single NV centers, we identified bare nanoposts containing single NV centers and compared their fluorescence lifetimes and single photon emission rates before and after Ag deposition. These optical characterizations were performed under ambient conditions using a home-built confocal microscope with a modest numerical aperture (NA~0.6) and long working distance (WD~4mm), which allowed us to optically access plasmonic nanostructures through the 500μm-thick bulk diamond material (Fig. 2(b)). A comparison between confocal scans taken on an array of r~65nm posts before (Fig. 2(d)) and after (Fig. 2(e)) Ag-deposition under identical experimental conditions indicated a more than two-fold enhancement of light emission by the nanoposts in the presence of Ag. The nature of photon emission was then further characterized by autocorrelation measurements to determine photon statistics and by triggered fluorescence decay measurements[27] to extract lifetime information.

We tested a total of 65 Ag-embedded nanoposts of radii 50nm and 65nm and explored the single photon nature of the emitter by measuring its autocorrelation function, $g^{(2)} = \langle I(t)*I(t+\tau) \rangle / \langle I(t) \rangle^2$, as a function of time delay, $\tau$, in a Hanbury Brown and Twiss experiment. Of these, five devices exhibited single photon character with $g^{(2)}(0)<0.5$ (Fig. 3, from the circled post in Fig. 2(d)). No background subtraction was performed in all data presented. In addition to the preservation of single photon character, we observed significant narrowing of the antibunching dips in comparison to those taken of nanoposts prior to Ag deposition under similar experimental conditions (Fig. 3(a)), which suggests an enhancement of the spontaneous emission rate due to the plasmonic cavity. Another 13 devices showed some degree of antibunching with $0.5 < g^{(2)}(0) < 1$, although many of these had previously exhibited single photon emission in the absence of Ag. The reductions in $g^{(2)}$ contrast in these cases may be due to plasmon-enhanced background fluorescence.

The SE rate enhancement was further investigated by time resolved photoluminescence measurements. Pulsed excitation was used to trigger an exponentially decaying fluorescence signal from which the lifetime information was extracted. The ensemble averaged lifetime of the densely populated NV centers in an unstructured area was measured to be 16.7±0.5ns, while the bare posts (without Ag) exhibited quenched NV emission leading to much longer lifetimes, with average values of 33.3±7.5ns for r~65nm and 38.3±7.3ns for r~50nm nanoposts containing single NV centers (Fig. 3(c)). The increase in lifetime with decreasing post radius can be explained by the reduced density of states for the radiative transition as a result of the nanostructuring[28]. FDTD simulations based on measured post dimensions further corroborate the extent of quenching observed in our experiments.

After Ag deposition, lifetimes of the NV centers were shortened by maximum factors of 6.6 for r~65nm posts and 4.8 for r~50nm posts and were observed to be as short as 5.2ns. A comparison of fluorescence decays measured for a representative device before and after Ag deposition is shown in Fig. 3(d). The spontaneous emission rate enhancement is accompanied by an increase in the count rates which can be observed in saturation measurements (Fig. 3(e)). For the representative bare nanopost shown in Fig. 3(e), the saturation intensity is 12.4kcps at a saturation power of 1.45mW. Meanwhile, our best diamond-plasmon device saturated at 101kcps at a similar pump power, showing an increase in photon emission that is comparable to the degree of enhancement in radiative decay. In comparison to single NV centers in the bulk measured using the same low N.A. setup (data not shown), the saturation intensities are roughly 2-3 times higher. In addition to TIR at the diamond-air interface, photon losses in the system can be attributed to the coupling of emitted photons to surface plasmons which propagate laterally across the chip, on diamond-silver surface, and could not be collected. Indeed, FDTD calculations show that only 4-5% of the emitted photons are captured by the collection optics, which suggests that further structure optimizations, such as the addition of gratings to increase the number of collected photons, might be necessary in the future[16].

Photoluminescence spectra taken of the Ag-embedded nanoposts consisted of the NV photoluminescence and the first and second order Raman signals of diamond (not shown). We normalized the background-subtracted PL spectrum of each metal-capped nanopost in the r~50nm by its bare counterpart (the latter taken by excitation through air to boost signal to background ratio) and observed resonance peaks (Fig. 4(a)) with quality factors ~10 over a central peak range between 697nm and 732nm, which are consistent with numerical modeling on our fabricated structures. The shift in resonance peaks can be explained by the sensitivity of the resonance to nanoscale deviations in the dimensions of different devices (Fig. 1(d)). Finally, the observed fluorescence was unequivocally assigned to NV center emission using room temperature, optically detected electron spin resonance (ESR) measurements[29]. We observed a characteristic NV-ESR spectrum with a dip in fluorescence occurring at an applied microwave frequency of 2.87MHz with a fluorescence contrast of 18.3%. The preservation of contrast compared to measurements on bulk NV centers suggests the viability of spin systems based on plasmon-enhanced NV emission and motivates further studies on the effect of the SE enhancement on the readout efficiency of the NV spin.

To illustrate the overall performance improvement, we have plotted the saturation intensity against the lifetime for a number of devices in the bare and Ag-capped cases. Fig. 4(d) shows a significant decrease in the lifetimes of the Ag-capped devices accompanied by an enhancement in photon emission. For our best device, the Purcell enhancement is 6.4 and 3.2 when compared to average lifetimes in the bare nanopost and unstructured diamond, respectively. These values are in good agreement with our modeling of the fabricated devices using dimensions obtained from SEM imaging. In addition to the geometry (i.e. truncated conical shapes of finished devices), the modeling takes into account the implantation depth, as well as the ambiguity in the polarization angle of the dipole moment arising from phonon mixing due to the [100] orientation of the diamond crystal plane[30]. Since the nature of this mixing is not well understood, we calculated the SE rate enhancement as a function of the dipole orientation and determined the spectrally averaged minima and maxima of the SE rate enhancements to be 2 and 3.6, showing a good convergence with experimental values. Moreover, the degree of SE enhancement is maximized when the dipole is placed at the center of the aperture and drops off as its axial position deviates from the field maximum and so larger Purcell enhancement (Fig. 1(d)) are anticipated with optimized implantation depth. Finally, modified device designs[16] will allow for collimated emission resulting in larger collection efficiencies of emitted photons.

The scalable method presented here provides controlled coupling in an emitter-plasmonic resonator system, for a large number of devices in parallel, and has yielded Purcell-enhanced single photon emission of NV centers. This geometry could serve as a basis and proof of principle for more complex diamond-plasmon structures[16] with potential applications in quantum information processing and nanoscale magnetic sensing based on spin or photonic qubits. Such diamond-plasmon devices can potentially provide stable and reliable systems for demonstrating enhanced zero-phonon line (ZPL) of the NV center, long-range coupling between qubits via surface plasmons, and improved optical readouts for single spin states. Finally, the scalability of our approach could also be applied towards coupling plasmon resonances to other optical transitions based on the NV or different color centers in diamond.

**Methods:**

**Fabrication:**

The arrays of nanoposts used in this experiment were fabricated in an electronic grade, type IIa CVD diamond sample that was implanted with nitrogen ions (at an energy of 14keV and a dose of $1.25 \times 10^{12}/cm^2$) and subsequently annealed under high vacuum ($<10^{-6}$Torr) at 750$^o$C for 2 hours to generate a layer of NV centers roughly 20nm below the diamond surface. The substrate was cleaned in a boiling 1:1:1 nitric, perchloric and sulphuric acid bath prior to resist spinning (XR electron beam resist, Dow Corning). Arrays of circular patterns (of radii 50nm to 65nm) were then defined using an EBL system (Elionix) at 100kV. After development of the resist (in Tetra-methyl ammonium hydroxide (TMAH, 25%)), the sample was subject to an ICP RIE oxygen dry etch to transfer the mask patterns onto the diamond substrate, resulting in ~180nm-tall nanoposts. The sample was then placed in hydrofluoric acid and the nitric, perchloric, and sulfuric acid bath to remove the residual mask as well as any contaminants from processing. A 500nm-thick Ag film was finally deposited on the nanoposts by electron beam evaporation (Denton).

**Characterization:**

Optical performance of the nanoposts before and after Ag deposition was tested in a home-built confocal microscope. For the autocorrelation and saturation measurements, a continuous-wave 532nm laser was used for excitation and focused through the diamond sample onto individual posts using an air objective (Olympus LUCPlanFLN 40x, NA 0.6), while both the incoming and collected signals were scanned by a steering mirror (Newport). Emitted light collected by the objective passed through a dichroic mirror and was then spectrally and spatially filtered using bandpass filters (650-800nm) and a single mode fiber (2x2 coupler, Thorlabs) before being sent to avalanche photodiodes (Perkin Elmer) for photodetection and measurement of photon statistics.  Spectral data were acquired using a grating spectrometer (Jobin Yvon iHR550, 76mm x 76mm monochromator with 150g/mm gratings).

Pulsed excitations used to trigger the decaying fluorescence signals were generated by passing ultrafast (~200fs) pulses at ~800nm from a Ti-Sapphire laser (Coherent Mira 800-F) through a photonic crystal fiber (Newport, SCG-800).  The resulting supercontinuum white light was spectrally filtered using bandpass filters (510-540nm, Semrock) to isolate green pulses. Prior to launching into the photonic crystal fiber, the fundamental repetition rate of the Ti:Sapphire pulse train (76MHz) was reduced to ~10.8 MHz using an electro-optic modulator (ConOptics, Model 350).  All time-correlated measurements were performed using a time-correlated-single-photon-counting module (PicoHarp).

ESR measurements were performed using a confocal microscope with a 0.8 NA objective. Microwaves were applied using a semirigid coaxial cable which was shorted with a 25µm-diameter Au bonding wire loop. The Au wire was approached from the bottom side of the sample in close proximity of the Ag film (30-40µm distance).  Bulk Ag has a skin-depth of 1.2µm at 2.8GHz and our 500nm Ag film is therefore largely transparent to the applied RF field. Additionally, the reduced conductivity of the evaporated Ag film further enhances its transparency to microwaves.  A Rohde Schwarz SMB 100A microwave generator was used and the signal was amplified with a 30dB gain in a Minicircuits ZHL-42W.

## **Acknowledgements**

We thank D. Twitchen and M. Markham from Element Six for providing diamond samples, and Chun Liang Yu and Philip Hemmer for helpful discussions.  T. M. B. acknowledges support from the NDSEG and NSF fellowships, and J. T. C. acknowledges support from the NSF graduate research fellowship.  Devices were fabricated in the Center for Nanoscale Systems (CNS) at Harvard.  This work was supported in part by Harvard's Nanoscale Science and Engineering Center (NSEC), NSF NIRT grant (ECCS-0708905), and by the DARPA QuEST program.  M. L. acknowledges support from the Sloan Foundation.

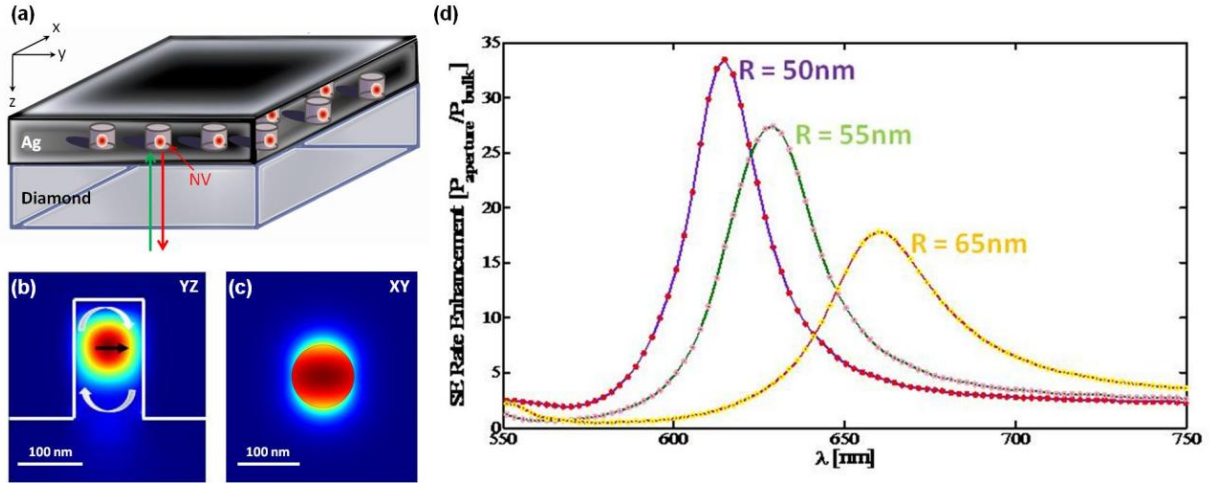

**Figure 1: A single photon source based on diamond-plasmon apertures.** **(a)** Three-dimensional schematic of the diamond-plasmonic system. The idealized structure for coupling to the NV emission consists of a diamond nanopost (with height~180nm and diameter~100nm) embedded in a 500nm-thick layer of silver. NV fluorescence is excited and collected through the bulk part of the diamond sample (green and red arrows, respectively). **(b)** Cross-sectional view of the structure along either the XZ (or YZ) plane, plotted with the longitudinal mode profile. The dipole is shown to be in-plane polarized and positioned in the center of the structure, where field density is maximized. As in a Fabry-Perot resonator, light reflects off the diamond-metal interfaces (curved arrows) and becomes tightly confined in the nanoposts. **(c)** Simulated lateral mode profile shows a near-uniform energy distribution of the plasmonic mode. **(d)** Simulated spontaneous emission enhancement as a function of wavelength for nanoposts with different radii, calculated with placing the dipole at the field maximum.

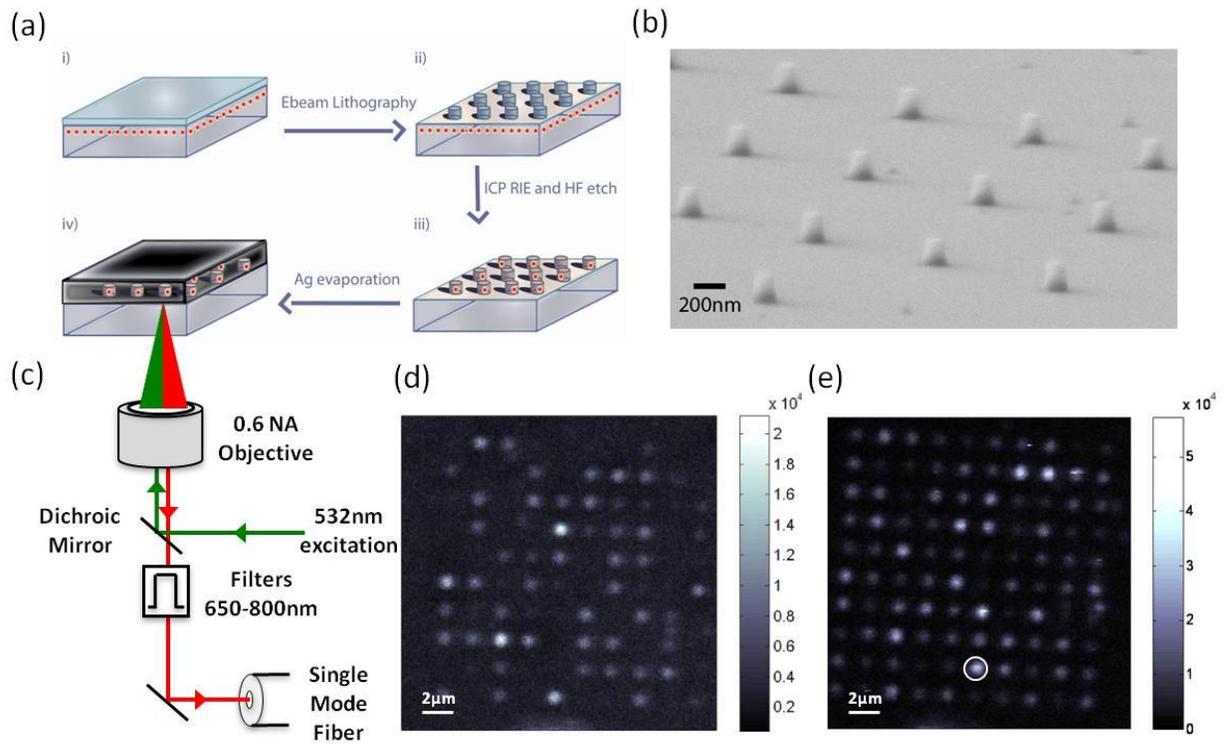

**Figure 2: Fabrication and optical microscopy of devices.** **(a)** Illustration of the fabrication procedure: (i) resist spinning on bulk diamond after nitrogen implantation and annealing, (ii) mask definition via e-beam lithography, (iii) pattern transfer to the diamond substrate in an oxygen-based RIE, and finally (iv) capping of the shallow implanted diamond posts in Ag (after resist removal). Optical characterization was performed both after step (iii) and (iv) on the same set of posts to measure the extent of plasmonic enhancement. **(b)** Scanning electron microscope image of a representative array of diamond posts after step (iii) in Fig. 2(a). **(c)** Schematic of the experimental configuration for sample characterization. **(d)-(e)** Confocal microscope scans of the same array of r~65nm posts before ((d)) and after ((e)) Ag-deposition under identical pump power. The spacing between adjacent posts is 2μm.

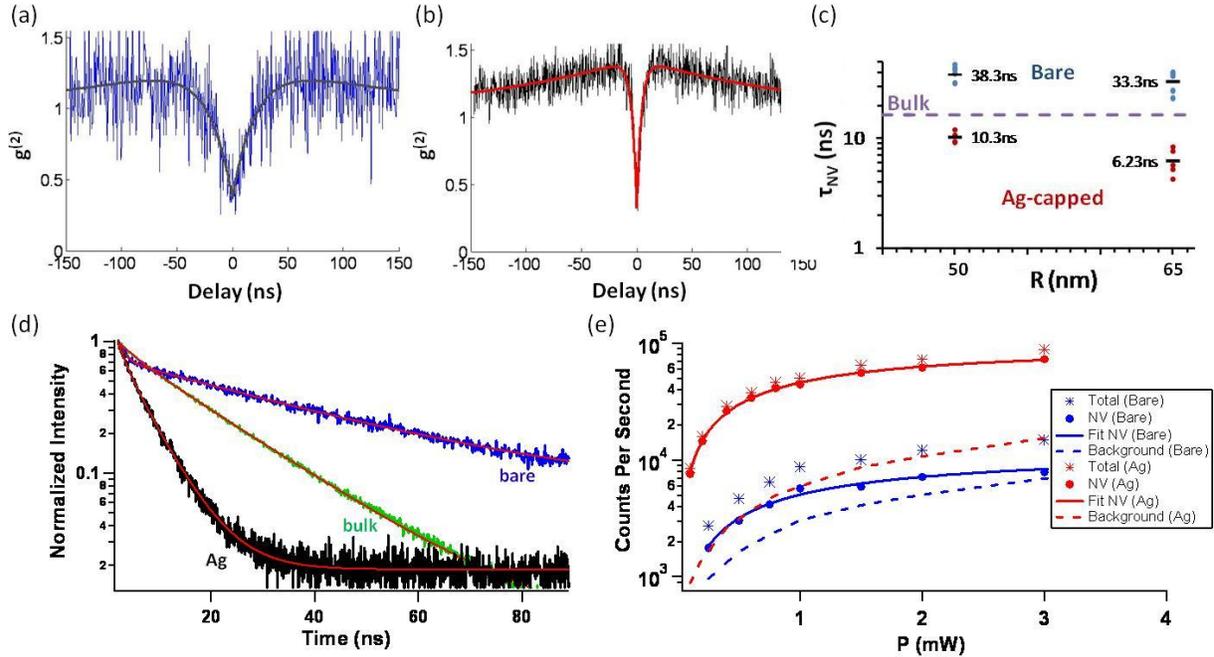

**Figure 3: Spontaneous emission enhancement of single NV centers.** Autocorrelation function for a **(a)** representative bare nanopost and **(b)** Ag-embedded device (circled in Fig. 2(d)) reveal strong antibunching at zero time delay indicating the emission of non-classical light. No background subtraction was performed. The gray and red curves represent fits to the g$^{(2)}$ function[1]. **(c)** NV lifetime as a function of radius before (total of 6 r~50nm and 5 r~65nm nanoposts) and after Ag deposition (total of 4 r~50nm and 5 r~65nm nanoposts). Blue represents bare while red denotes Ag-capped nanoposts. The dotted line represents the ensemble averaged lifetimes measured in an unstructured area of the sample, while the black lines indicate the average value of each set. The quenching of light emission in the bare nanoposts is due to the reduced density of states in the nanostructures. **(d)** Normalized fluorescence decays for the same r~65nm nanopost, containing a single NV center, before and after Ag-deposition and for an ensemble population of NVs in the bulk region, along with fits to a multi-exponential model (shown in red). The fits yielded time constants for fast-decaying background fluorescence (<2ns) as well as a slower NV photoluminescence (bare: 37.17±0.7ns; Ag-embedded: 5.65±0.08ns; bulk: 16.7±0.08ns). **(e)** Saturation curves for representative single-photon-emitting nanoposts in the bare (blue) and Ag-embedded (red) cases. In each scenario, the total count rates are represented by the asterisks, while the background contributions (as measured from an empty post not containing any NV centers) are denoted by the dotted line. Subtraction of the background from the total yields the NV emission (dots), which can be fitted to the saturation model (solid lines). The fitted saturation intensities and powers are 1.01±0.02 x10$^5$ cps and 1.18±0.06 mW for the Ag-embedded post, and 1.24±0.09 x10$^4$cps and 1.45±0.23 mW for the bare case.

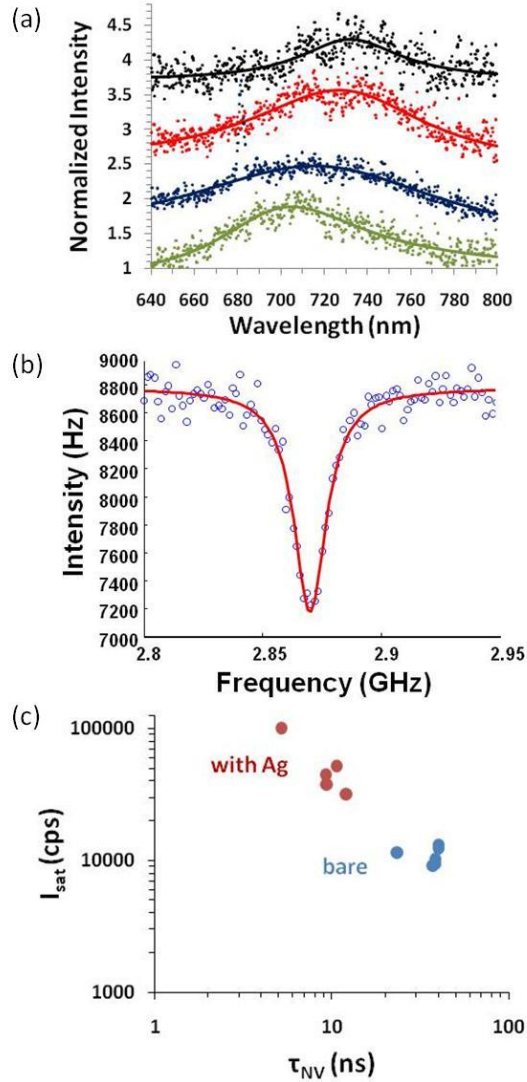

**Figure 4: Comparison of photoluminescence spectra and device performances.** **(a)** Normalized spectra for four different devices in the r~50nm array obtained by dividing the background-subtracted PL signals from the Ag-embedded nanoposts over those of the bare nanoposts (taken by excitation and collection through air). Each solid line represents the Fano fit to the raw data (dots). The quality factors corresponding to the fits are between 5 (blue) and 13 (black), while the resonance peaks range from 697nm and 732nm. A constant offset has been applied to the normalized intensity for clarity. **(b)** Optically detected magnetic resonance (ODMR) spectrum of a plasmon-enhanced NV center, measured by tuning the microwave over the NV center splitting between the $m_s$=0 and $m_s$=±1 ground state levels without an external magnetic field. The spectrum reveals a characteristic dip at 2.87GHz and a contrast of 18.3%. **(c)** Performance plot of saturation intensity as a function of fluorescence lifetime for the bare and Ag-capped devices we tested for this study, showing that the plasmonic enhancement provided by the geometry has led to shorter lifetimes and correspondingly higher count rates.